# Effect of inorganic cation mixing on hybrid halide perovskites using density functional theory for application in single and multi junction photovoltaics

**Manaswita Kar***
University of Southampton,
M.Kar@soton.ac.uk

### Abstract

This work is aimed to study mixed-cation halide perovskites of the form AA'BX3 (A = methylammonium $CH_3NH_3$; A' = K, Rb, or Cs; B = Pb, Sn, or Ge; X = Cl, Br, or I), using first principles Density Functional Theory (DFT) formalism in order to find their potential applications in perovskite based photovoltaics. This is an addition to the two previous studies performed by the author [1-2] on high-throughput screening of hybrid halide and inorganic halide perovskites. This work involves mixing certain amounts of the inorganic cation to the organic site and study the electronic structure of the resultant. 63 perovskites have been simulated in their calculated stable phases and their band gaps have been predicted. From the calculated band gaps, 37 perovskites are predicted to be suitable for single junction as well as multi junction solar cell application, 7 are not suitable for either single or multi junction solar cells, and the rest 21 are suitable for multi junction but not for single junction solar cells. The study also shows interesting transition in the nature of the band gaps caused at about 50% of inorganic cation mixing.

### Introduction

Ever since the ground-breaking record of power conversion efficiency (PCE) of 9.1%, [3] hybrid halide perovskites have dominated the solar cell research regime as promising light-harvesting materials for photovoltaic applications. Within a short span of research time, the efficiencies of these perovskites soared to 22.1%, attracting the attention of the photovoltaic community. [4-5] The incredible success attained by these perovskites is owed to their superior optical and electronic properties, for example, high light absorption coefficient, high charge carrier mobility, and low exciton binding energy, etc., to name a few. [6-8] These properties along with the low cost of these materials due to utilisation of earth abundant elements have made the perovskites surge the photovoltaics market in the recent years. [9-11] These unique features owe direct contribution to the competition of the perovskite solar cells (PSCs) over the previous generation PV technologies that are based on commercial silicon (20%), GaAs (18%), cadmium telluride (CdTe-19%), and copper indium gallium selenide/sulphide (CIGS, 19%). Here, the figures in the parentheses are the indicators of the maximum PCEs of the commercial devices for each of the mentioned technologies. [12-18]

However, the commercialisation of the most efficient perovskite ($MAPbI_3$) faces severe challenges, primarily due to two reasons: lack of stability and the presence of lead that might have severe implications on people and the environment. The latter issue has been addressed by several groups either by on-device lead sequestration [19] or by replacing the lead by other metals, such as tin, germanium, etc. using high-throughput screening. [20-30] Nonetheless, the commercial viability of these perovskite materials are hindered to a large extent by the fact that they are unstable and degrade easily under prolonged exposure to moisture. [31-36]

Doping impurities on different sites has been found to be effective in order to improve the performances of these perovskite materials. Incorporating small amounts of K, Cs, and Rb in hybrid perovskites have proved to improve not only their phase stability, but also the mixed-cation perovskites have consistently exhibited higher open-circuit voltage, short-circuit current, fill factor, and long term stability towards light and moisture. [37-43]

In this work, mixed-cation halide perovskites are explored by doping the methylammonium site with 3 inorganic cations; namely K, Rb, and Cs in varied proportions. The 0K stable phases of the mixed perovskites are studied by calculating the Goldschmidt tolerance factor using the ionic radii of the constituents.[44] As a next step, the energy band gaps of the perovskites are predicted using DFT formalism in the phases predicted previously [45-50]. It is seen that the predicted band gaps of the mixed cation perovskites are suitable for either single junction or tandem solar cell application as set by the Shockley-Queisser limit.[51]

### Computational Details

All the calculations are done using Density Functional Theory (DFT) within the Vienna Ab-initio Software Package (VASP) [52-53]. The initial geometries of the perovskites in the respective stable phases are obtained from Materials Project software. The geometries of the perovskites and their band gaps are calculated using the Perdew-Burke-Ernzerhof (PBE) exchange-correlation functional [54-55]. An energy cut-off of 400 eV is used for the geometry optimisations and the band gap calculations. A k-point grid of 3x3x3 is used for geometry optimisation and 6x6x6 for the band gap calculations. Plane-wave ultrasoft pseudo-potentials have been used for all the elements constituting the study [56-57]. All parameters have been tested for computational convergence.

## Results and Discussions

### a) Phase stability

The phase stability of the perovskite materials is an important aspect to consider in order to find their suitable applications. However, predicting the most stable structure for complex materials such as perovskites is an ongoing challenge in chemistry. Even in cases where modern computational methods have significantly improved the chances to successfully predict stability of structures, these methods are time-consuming and are not suitable for all compounds. For complex materials such as hybrid perovskites, the computational approaches fail to predict the rotation of the organic cation within the perovskite structure. Empirical methods based on simple geometric approaches have proved to be a better alternative for predicting the stability of structures over years. The Goldschmidt tolerance factor is one such empirical method that is used to predict the room temperature phase stability of the perovskites [44]. The Goldschmidt tolerance factor (t) is calculated as:

$$\boldsymbol{t} = \frac{\boldsymbol{r_A}+\boldsymbol{r_X}}{\sqrt{2}(\boldsymbol{r_B}+\boldsymbol{r_X})} \qquad (1)$$

Here, $r_A$, $r_B$, and $r_X$ are the radii of the A, B, and X ions. t=1 indicates a perfect, cubic perovskite. A shift of the value from t=1 indicates a distortion in the structure. A value less than 1 implies an orthorhombic structure, whereas t > 1 implies a distorted triclinic/monoclinic structure. For mixed cation perovskites, the Goldschmidt tolerance factor is calculated by considering the ratio of individual organic and inorganic cations. For instance, the GTF of a mixed halide perovskite with y amount of MA and (1-y) amount of inorganic cation is calculated as:

$$\boldsymbol{t} = \frac{(\boldsymbol{1}-\boldsymbol{y})\boldsymbol{r_{MA}}+(\boldsymbol{y})\boldsymbol{r_{in}}+\boldsymbol{r_X}}{\sqrt{2}(\boldsymbol{r_B}+\boldsymbol{r_X})} \qquad (2)$$

Here, $r_{MA}$ is the radius of the methylammonium cation, $r_{in}$ the radius of the corresponding inorganic cation, $r_B$ the radius of the metal ion and $r_X$ is the radius of the halide ion. The radius of methylammonium cation was initially calculated by Amat et al., [58] by calculating the volume inside a contour of 0.001 electrons/Bohr$^3$ density. The value obtained for the radius was 2.70 Å. Kieslich et al considered a hard sphere model within which the cation rotates freely about its center of mass [59]. This method yielded a radius value of 2.17 Å. The bound on the radius of the methylammonium cation is set by considering that $MAPbBr_3$ and $MAPbCl_3$ have a perfect cubic structure, thus having t between 0.9 and 1.0. [60-63] Considering the ionic radii of Pb, Br, and Cl to be 1.19 Å, 1.96 Å, and 1.81 Å respectively, in order to maintain t between 0.9 and 1.0, the radius of MA should be between 2.04 Å and 2.50 Å. Hence, the radius is considered to be 2.17 Å. The radii of the inorganic cations, the metal ions and the halide ions are shown in table 1. The tolerance factor of the simulated mixed cation perovskites and their corresponding stable phases are shown in table 2. From this table, it is seen that all the pure hybrid perovskites assume a cubic crystal structure based on their t values. However, by mixing certain amounts of inorganic cations, the crystal structure changes from cubic to orthorhombic. This is due to the change in the cations and their concentration in the perovskite materials.

| Ion | Ionic radius (Å) |
|---|---|
| CH3NH3+ | 2.17 |
| K+ | 1.38 |
| Cs+ | 1.67 |
| Rb+ | 1.52 |
| Pb2+ | 1.19 |
| Sn2+ | 1.18 |
| Ge2+ | 1.22 |
| I- | 2.20 |
| Br- | 1.96 |
| Cl- | 1.84 |

**Table 1: Ionic radii of the constituents of the mixed perovskites used for Goldschmidt tolerance factor calculation.**

| A | B | X | y | GTF (t) | Predicted stable phase |
|---|---|---|---|---|---|
| K | Pb | Br | 0 | 0.92 | cubic |
| K | Pb | Br | 0.25 | 0.88 | orthorhombic |
| K | Pb | Br | 0.5 | 0.839 | orthorhombic |
| K | Pb | Cl | 0 | 0.93 | cubic |
| K | Pb | Cl | 0.25 | 0.887 | orthorhombic |
| K | Pb | Cl | 0.5 | 0.84 | orthorhombic |
| K | Pb | I | 0 | 0.91 | cubic |
| K | Pb | I | 0.25 | 0.87 | orthorhombic |
| K | Pb | I | 0.5 | 0.85 | orthorhombic |
| K | Ge | Br | 0 | 0.92 | cubic |
| K | Ge | Br | 0.25 | 0.89 | orthorhombic |
| K | Ge | Br | 0.5 | 0.86 | orthorhombic |
| K | Ge | Cl | 0 | 0.95 | cubic |
| K | Ge | Cl | 0.25 | 0.88 | orthorhombic |
| K | Ge | Cl | 0.5 | 0.86 | orthorhombic |
| K | Ge | I | 0 | 0.9 | cubic |
| K | Ge | I | 0.25 | 0.88 | orthorhombic |
| K | Ge | I | 0.5 | 0.85 | orthorhombic |
| K | Sn | Br | 0 | 0.93 | cubic |
| K | Sn | Br | 0.25 | 0.86 | orthorhombic |
| K | Sn | Br | 0.5 | 0.83 | orthorhombic |
| K | Sn | Cl | 0 | 0.92 | cubic |
| K | Sn | Cl | 0.25 | 0.88 | orthorhombic |
| K | Sn | Cl | 0.5 | 0.86 | orthorhombic |
| K | Sn | I | 0 | 0.93 | cubic |
| K | Sn | I | 0.25 | 0.87 | orthorhombic |
| K | Sn | I | 0.5 | 0.83 | orthorhombic |
| Rb | Pb | Br | 0 | 0.95 | cubic |
| Rb | Pb | Br | 0.25 | 0.87 | orthorhombic |
| Rb | Pb | Br | 0.5 | 0.83 | orthorhombic |
| Rb | Pb | Cl | 0 | 0.97 | cubic |
| Rb | Pb | Cl | 0.25 | 0.86 | orthorhombic |
| Rb | Pb | Cl | 0.5 | 0.82 | orthorhombic |
| Rb | Pb | I | 0 | 0.9 | cubic |
| Rb | Pb | I | 0.25 | 0.84 | orthorhombic |
| Rb | Pb | I | 0.5 | 0.82 | orthorhombic |
| Rb | Ge | Br | 0 | 0.95 | cubic |
| Rb | Ge | Br | 0.25 | 0.87 | orthorhombic |
| Rb | Ge | Br | 0.5 | 0.83 | orthorhombic |
| Rb | Ge | Cl | 0 | 0.97 | cubic |
| Rb | Ge | Cl | 0.25 | 0.86 | orthorhombic |
| Rb | Ge | Cl | 0.5 | 0.82 | orthorhombic |
| Rb | Ge | I | 0 | 0.9 | cubic |
| Rb | Ge | I | 0.25 | 0.84 | orthorhombic |
| Rb | Ge | I | 0.5 | 0.82 | orthorhombic |
| Rb | Sn | Br | 0 | 0.95 | cubic |
| Rb | Sn | Br | 0.25 | 0.87 | orthorhombic |
| Rb | Sn | Br | 0.5 | 0.83 | orthorhombic |
| Rb | Sn | Cl | 0 | 0.97 | cubic |
| Rb | Sn | Cl | 0.25 | 0.86 | orthorhombic |
| Rb | Sn | Cl | 0.5 | 0.82 | orthorhombic |
| Rb | Sn | I | 0 | 0.9 | cubic |
| Rb | Sn | I | 0.25 | 0.84 | orthorhombic |
| Rb | Sn | I | 0.5 | 0.82 | orthorhombic |
| Cs | Pb | Br | 0 | 0.95 | cubic |
| Cs | Pb | Br | 0.25 | 0.87 | orthorhombic |
| Cs | Pb | Br | 0.5 | 0.83 | orthorhombic |
| Cs | Pb | Cl | 0 | 0.97 | cubic |
| Cs | Pb | Cl | 0.25 | 0.86 | orthorhombic |
| Cs | Pb | Cl | 0.5 | 0.82 | orthorhombic |
| Cs | Pb | I | 0 | 0.9 | cubic |
| Cs | Pb | I | 0.25 | 0.84 | orthorhombic |
| Cs | Pb | I | 0.5 | 0.82 | orthorhombic |
| Cs | Ge | Br | 0 | 0.95 | cubic |
| Cs | Ge | Br | 0.25 | 0.87 | orthorhombic |
| Cs | Ge | Br | 0.5 | 0.83 | orthorhombic |
| Cs | Ge | Cl | 0 | 0.97 | cubic |
| Cs | Ge | Cl | 0.25 | 0.86 | orthorhombic |
| Cs | Ge | Cl | 0.5 | 0.82 | orthorhombic |
| Cs | Ge | I | 0 | 0.9 | cubic |
| Cs | Ge | I | 0.25 | 0.84 | orthorhombic |
| Cs | Ge | I | 0.5 | 0.82 | orthorhombic |
| Cs | Sn | Br | 0 | 0.95 | cubic |
| Cs | Sn | Br | 0.25 | 0.87 | orthorhombic |
| Cs | Sn | Br | 0.5 | 0.83 | orthorhombic |
| Cs | Sn | Cl | 0 | 0.97 | cubic |
| Cs | Sn | Cl | 0.25 | 0.86 | orthorhombic |
| Cs | Sn | Cl | 0.5 | 0.82 | orthorhombic |
| Cs | Sn | I | 0 | 0.9 | cubic |
| Cs | Sn | I | 0.25 | 0.84 | orthorhombic |
| Cs | Sn | I | 0.5 | 0.82 | orthorhombic |

**Table 2: Predicted room temperature stable phases of the simulated perovskites using Goldschmidt tolerance factor.**

### b) Energy band gaps

The energy band gaps of the mixed perovskites and their nature have been predicted in this study. These factors are an indicator as to whether these materials are suitable for photovoltaic applications. The Shockley-Queisser limit determines that the optimal band gap for a single p-n junction solar cell device of highest efficiency should be 1.3 eV [50]. In case of tandem solar cells, according to Alexis De Vos in 1980,[64] the band gaps of the 2 individual perovskites should be around 1.0 eV and 1.9 eV. Owing to the computational inaccuracies, a margin of 0.5 eV has been kept towards the prediction of these band gaps in this study. Thus, if the predicted band gap lies between 0.8 eV and 1.8eV, then the perovskite is suitable for single junction PV application; and if the same lies between 0.5 eV and 2.5 eV, then those are suitable for two-layer tandem solar cells. However, it must be noted that the band gaps predicted are DFT band gaps and can be quite different from the experimental band gaps. [65-66] However, there is no significant experimental literature available to compare the particular compositions of the mixed halide perovskites that are simulated in this study. [66] The band gaps predicted in this study aim to find suitable applications of these perovskites as single junction or tandem solar cells as shown in table 3. The predicted band gaps also show that there is a change in the nature of the band gaps from indirect to direct by mixing inorganic cations to the hybrid perovskites. The magnitude and the nature of the band gaps of the perovskites depend on the composition of the A, B and X ions within the perovskites, and the different initial conditions set by the computational parameters. In general, indirect band gaps in perovskites result in long carrier lifetimes of these materials as the thermalised carriers are forbidden from recombining via direct transition. Rashba splitting of the conduction band results in an indirect transition, causing the band to split due to the local electric field generated on the p-orbitals of the B-site metal atoms that constitute most of the conduction band [67]. Nevertheless, semiconductors with indirect band gaps are also used in solar cell applications [68]. Mixing certain amounts of inorganic cations result in the transition of the band gaps from indirect to direct as is seen in the table and is thus expected to solve the long term stability issue in these hybrid perovskites.

| **A** | **B** | **X** | **y** | **Predicted band gap in the stable phase (eV)** | **Nature** | **Magnitude suitable for single junction PV application** | **Magnitude suitable for double layer tandem PV application** |
|---|---|---|---|---|---|---|---|
| K | Pb | Br | 0 | 2.320 | Indirect | No | Yes |
| K | Pb | Br | 0.25 | 2.239 | Indirect | No | Yes |
| K | Pb | Br | 0.5 | 2.370 | Direct | No | Yes |
| K | Pb | Cl | 0 | 2.720 | Direct | No | No |
| K | Pb | Cl | 0.25 | 2.610 | Indirect | No | No |
| K | Pb | Cl | 0.5 | 2.888 | Direct | No | No |
| K | Pb | I | 0 | 1.547 | Direct | Yes | Yes |
| K | Pb | I | 0.25 | 1.494 | Direct | Yes | Yes |
| K | Pb | I | 0.5 | 1.552 | Indirect | Yes | Yes |
| K | Ge | Br | 0 | 1.647 | Direct | Yes | Yes |
| K | Ge | Br | 0.25 | 1.796 | Indirect | Yes | Yes |
| K | Ge | Br | 0.5 | 1.880 | Direct | Yes | Yes |
| K | Ge | Cl | 0 | 2.296 | Indirect | No | Yes |
| K | Ge | Cl | 0.25 | 2.180 | Indirect | No | Yes |
| K | Ge | Cl | 0.5 | 2.303 | Direct | No | Yes |
| K | Ge | I | 0 | 1.313 | Indirect | Yes | Yes |
| K | Ge | I | 0.25 | 1.341 | Direct | Yes | Yes |
| K | Ge | I | 0.5 | 1.310 | Direct | Yes | Yes |
| K | Sn | Br | 0 | 1.415 | Direct | Yes | Yes |
| K | Sn | Br | 0.25 | 1.770 | Indirect | Yes | Yes |
| K | Sn | Br | 0.5 | 1.892 | Direct | No | Yes |
| K | Sn | Cl | 0 | 2.003 | Indirect | No | Yes |
| K | Sn | Cl | 0.25 | 2.217 | Indirect | No | Yes |
| K | Sn | Cl | 0.5 | 2.283 | Direct | No | Yes |
| K | Sn | I | 0 | 0.890 | Indirect | Yes | Yes |
| K | Sn | I | 0.25 | 1.312 | Indirect | No | Yes |
| K | Sn | I | 0.5 | 1.462 | Direct | No | Yes |
| Rb | Pb | Br | 0 | 2.320 | Indirect | No | Yes |
| Rb | Pb | Br | 0.25 | 2.219 | Indirect | No | Yes |
| Rb | Pb | Br | 0.5 | 2.371 | Direct | No | Yes |
| Rb | Pb | Cl | 0 | 2.320 | Indirect | No | Yes |
| Rb | Pb | Cl | 0.25 | 2.210 | Indirect | No | Yes |
| Rb | Pb | Cl | 0.5 | 2.787 | Direct | No | No |
| Rb | Pb | I | 0 | 1.547 | Indirect | No | Yes |
| Rb | Pb | I | 0.25 | 1.527 | Indirect | No | Yes |
| Rb | Pb | I | 0.5 | 1.967 | Direct | No | Yes |
| Rb | Ge | Br | 0 | 1.647 | Indirect | No | Yes |
| Rb | Ge | Br | 0.25 | 1.797 | Indirect | No | Yes |
| Rb | Ge | Br | 0.5 | 1.881 | Direct | No | Yes |
| Rb | Ge | Cl | 0 | 2.296 | Indirect | No | Yes |
| Rb | Ge | Cl | 0.25 | 2.186 | Indirect | No | Yes |
| Rb | Ge | Cl | 0.5 | 2.302 | Direct | No | Yes |
| Rb | Ge | I | 0 | 1.313 | Indirect | Yes | Yes |
| Rb | Ge | I | 0.25 | 1.259 | Indirect | Yes | Yes |
| Rb | Ge | I | 0.5 | 1.315 | Direct | Yes | Yes |
| Rb | Sn | Br | 0 | 1.415 | Direct | Yes | Yes |
| Rb | Sn | Br | 0.25 | 1.775 | Indirect | Yes | Yes |
| Rb | Sn | Br | 0.5 | 1.986 | Direct | Yes | Yes |
| Rb | Sn | Cl | 0 | 2.001 | Indirect | No | Yes |
| Rb | Sn | Cl | 0.25 | 2.237 | Indirect | No | Yes |
| Rb | Sn | Cl | 0.5 | 2.270 | Direct | No | Yes |
| Rb | Sn | I | 0 | 0.890 | Direct | Yes | No |
| Rb | Sn | I | 0.25 | 1.301 | Indirect | Yes | Yes |
| Rb | Sn | I | 0.5 | 1.466 | Direct | Yes | Yes |
| Cs | Pb | Br | 0 | 2.320 | Indirect | No | Yes |
| Cs | Pb | Br | 0.25 | 2.261 | Indirect | No | Yes |
| Cs | Pb | Br | 0.5 | 2.325 | Direct | No | Yes |
| Cs | Pb | Cl | 0 | 2.720 | Indirect | No | No |
| Cs | Pb | Cl | 0.25 | 2.863 | Indirect | No | No |
| Cs | Pb | Cl | 0.5 | 2.944 | Direct | No | No |
| Cs | Pb | I | 0 | 1.547 | Indirect | Yes | Yes |
| Cs | Pb | I | 0.25 | 1.517 | Indirect | Yes | Yes |
| Cs | Pb | I | 0.5 | 1.564 | Direct | Yes | Yes |
| Cs | Ge | Br | 0 | 1.647 | Indirect | Yes | Yes |
| Cs | Ge | Br | 0.25 | 1.802 | Indirect | Yes | Yes |
| Cs | Ge | Br | 0.5 | 1.821 | Direct | No | Yes |
| Cs | Ge | Cl | 0 | 2.296 | Indirect | No | Yes |
| Cs | Ge | Cl | 0.25 | 2.191 | Indirect | No | Yes |
| Cs | Ge | Cl | 0.5 | 2.322 | Direct | No | Yes |
| Cs | Ge | I | 0 | 1.313 | Indirect | Yes | Yes |
| Cs | Ge | I | 0.25 | 1.335 | Indirect | Yes | Yes |
| Cs | Ge | I | 0.5 | 1.356 | Direct | Yes | Yes |
| Cs | Sn | Br | 0 | 1.415 | Indirect | Yes | Yes |
| Cs | Sn | Br | 0.25 | 1.540 | Indirect | Yes | Yes |
| Cs | Sn | Br | 0.5 | 1.592 | Direct | Yes | Yes |
| Cs | Sn | Cl | 0 | 2.003 | Indirect | No | Yes |
| Cs | Sn | Cl | 0.25 | 2.122 | Indirect | No | Yes |
| Cs | Sn | Cl | 0.5 | 2.171 | Direct | No | Yes |
| Cs | Sn | I | 0 | 0.890 | Indirect | Yes | Yes |
| Cs | Sn | I | 0.25 | 1.671 | Indirect | Yes | Yes |
| Cs | Sn | I | 0.5 | 1.712 | Direct | Yes | Yes |

**Table 3: Predicted energy band gaps of the simulated perovskites using DFT-PBE formalism.**

## Conclusions

From this study, a number of mixed-cation perovskites have been identified that are considered suitable for applications in single junction and tandem solar cells. It is seen that perovskites containing I at the halide site has significantly lower band gaps than the ones with Br and Cl. Therefore, the iodine-based perovskites have more potential to be used as both single junction as well as in tandem solar cell applications. The Br and Cl based perovskites typically have higher values for the predicted band gaps and hence are more suitable for tandem solar cell applications. 37 mixed cation perovskites have been predicted to be suitable for both single and multi junction solar cells, 21 perovskites are suitable for multi junction solar cells but not for single junction, and 7 are not suitable for either single junction or multi-junction solar cells. In general, the mixed-cation halide perovskites have several advantages over pure hybrid halide perovskites as already mentioned previously. However, there is very little experimental literature available on these materials. Experimentally, there has been evidence of the synthesis of few mixed cation perovskites [69-70] It is thus expected that more of these mixed cation perovskites be synthesised and studied in order to exploit their full potential in the photovoltaics research community.

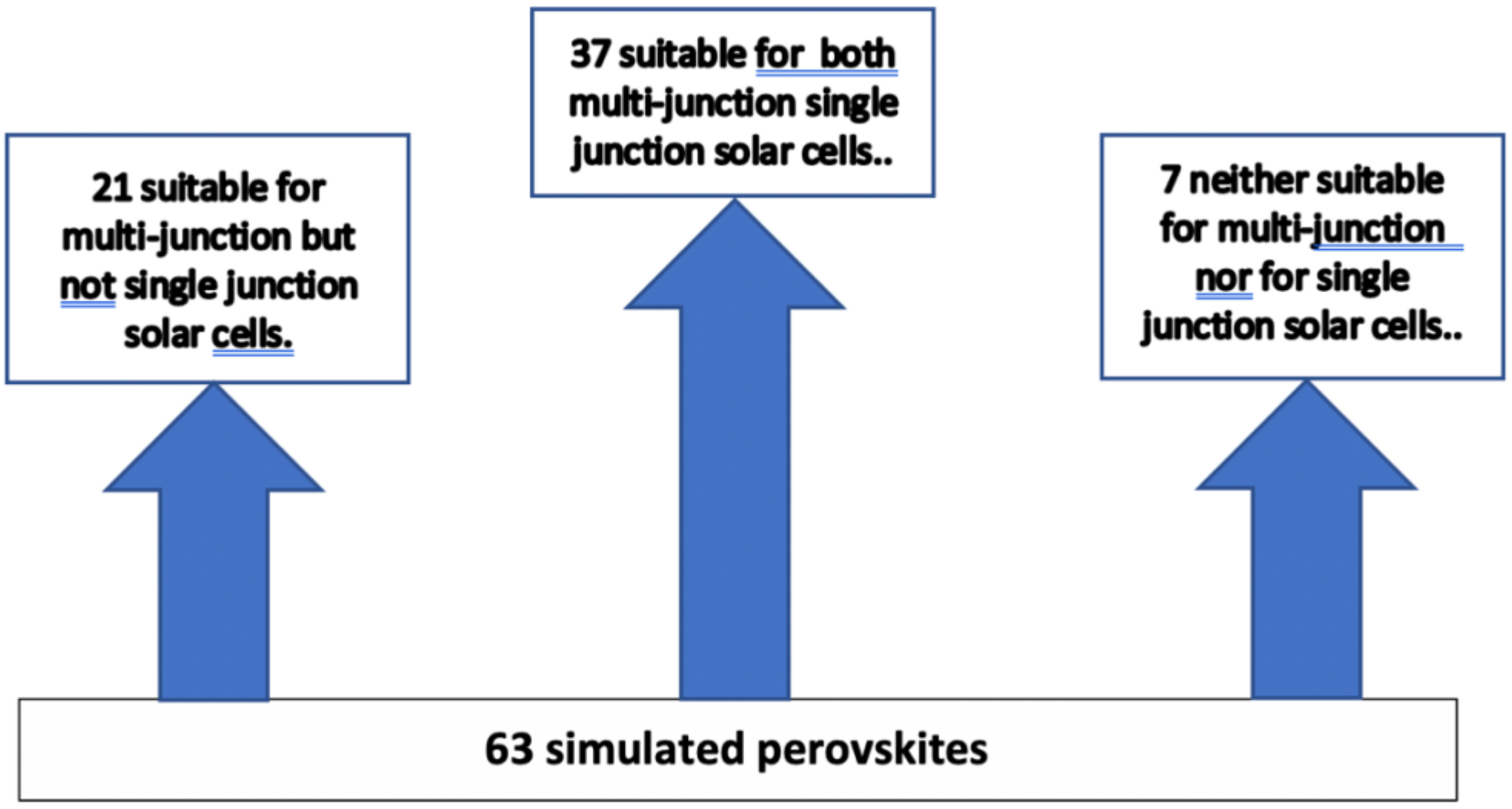

**Figure 1: Graphical depiction of the perovskite simulation used for solar cell applications.**

## Data Availability

All the data for the calculations used in this study are made available in the NOMAD repository.

## Acknowledgements

The author thanks the HyPerCells graduate school, Prof Dr Thomas Körzdörfer within the University of Potsdam, for giving the opportunity, guidance, as well as the necessary funding to support the project. The necessary computation time has been provided by the supercomputing cluster IRIDIS5 within the University of Southampton.